# Interrogating the legend of Einstein's "biggest blunder"

Cormac O'Raifeartaigh[§] and Simon Mitton[§§]

Full addresses of authors

[§]*School of Science and Computing, Waterford Institute of Technology, Cork Road, Waterford, Ireland.*

[§§]*St Edmund's College, University of Cambridge, Cambridge CB3 0BN, United Kingdom.*

Author for correspondence: coraifeartaigh@wit.ie

Short biographies of authors

[§] Cormac O'Raifeartaigh lectures in physics at Waterford Institute of Technology, Ireland and is Visiting Associate Professor of Physics at University College Dublin.

[§§] Simon Mitton is a historian of science and Life Fellow of St Edmund's College, University of Cambridge

**Abstract**

It is well known that, following the emergence of the first evidence for an expanding universe, Albert Einstein banished the cosmological constant term from his cosmology. Indeed, he is reputed to have labelled the term, originally introduced to the field equations of general relativity in 1917 in order to predict a static universe, his "biggest blunder". However serious doubts about this reported statement have been raised in recent years. In this paper, we interrogate the legend of Einstein's "biggest blunder" statement in the context of our recent studies of Einstein's cosmology in his later years. We find that the remark is highly compatible with Einstein's cosmic models of the 1930s, with his later writings on cosmology, and with independent reports by at least three physicists. We conclude that there is little doubt that Einstein came to view the introduction of the cosmological constant term a serious error and that it is very likely that he labelled the term his "biggest blunder" on at least one occasion. This finding may be of some relevance for those theoreticians today who seek to describe the recently-discovered acceleration in cosmic expansion without the use of a cosmological constant term.

*Key words:* biggest blunder; cosmological constant; general theory of relativity; relativistic cosmology; static universe; Albert Einstein; George Gamow; Hubble's law; expanding universe; accelerated expansion

# 1. Introduction

Does it matter whether or not Einstein once described his introduction of the cosmological constant term to the field equations as his "biggest blunder"? We recently decided to investigate this story for a number of reasons. In the first instance, if it can be shown that the report is false, the legend constitutes an important example of the dangers of taking an apocryphal story at face value. On the other hand, if it can be shown that the report is true, it casts useful light on Einstein's cosmology in his later years and may be of relevance to today's cosmologists. Finally, the story's strong association with the enigmatic physicist George Gamow prompts some reflections on one of the most intriguing and controversial figures of 20[th] century physics.

## 2. Einstein's introduction of the cosmological constant term

In 1917, Einstein attempted the first relativistic model of the universe[1] as an important test for his newly-minted general theory of relativity[2]. As he remarked to the Dutch astronomer Willem de Sitter: *"For me… it was a burning question whether the relativity concept can be followed through to the finish, or whether it leads to contradictions"*[3]. However, assuming a static, uniform distribution of matter and a cosmos of closed spatial curvature (due to his desire to render his model compatible with his understanding of Mach's Principle[4]), Einstein soon found that the covariant field equations of relativity

$$G_{\mu\nu} = -\kappa \left( T_{\mu\nu} - \tfrac{1}{2} g_{\mu\nu} T \right) \qquad (1)$$

gave a null solution for the case of the universe as a whole. (Here $G_{\mu\nu}$ is a four-dimensional tensor representing the curvature of space-time with elements $g_{\mu\nu}$ , $T_{\mu\nu}$ is a tensor representing energy and momentum, $T$ is a scalar and $\kappa$ is the Einstein constant $8\pi G/c^2$). Einstein's answer was to modify the field equations by introducing a new term according to

$$G_{\mu\nu} - \lambda g_{\mu\nu} = -\kappa \left( T_{\mu\nu} - \tfrac{1}{2} g_{\mu\nu} T \right) \qquad (2)$$

where λ was a universal constant known as the *cosmological constant*. Einstein then showed that the modified field equations (2) have the solution

$$\lambda = \frac{\kappa\rho}{2} = \frac{1}{R^2} \qquad (3)$$

where $\rho$ and $R$ represent the mean density of matter and the radius of the cosmos respectively[1]. Thus, Einstein's 1917 model of the cosmos gave an apparently satisfactory relation between the size of the universe and the amount of matter it contained. Indeed, in his correspondence around this time, Einstein attempted a rough estimate of the size of the universe (and thus of the cosmological constant) using estimates of the density of matter in the Milky Way, although he later realised that such calculations were unreliable[5].

However, there is little doubt that an interpretation of the physics underlying the cosmological constant term posed a challenge for Einstein from the outset. As we have previously pointed out[5], no discussion of the physics underlying the term is presented in Einstein's 1917 paper and his contemporaneous correspondence indicates that he viewed $\lambda$ as an uncomfortable mathematical necessity. For example, in March 1917 Einstein remarked to Felix Klein: "*The new version of the theory means, formally, a complication of the foundations and will probably be looked upon by almost all our colleagues as an interesting, though mischievous and superfluous stunt… But I see the matter as a necessary addition, without which neither inertia nor geometry are truly relative*"[6]. A month later, he made a rather prescient remark to the noted Dutch theorist and astronomer Willem de Sitter: "*The general theory of relativity allows the addition of the term $\lambda g_{\mu\nu}$ in the field equations. One day, our actual* *knowledge of the composition of the fixed-star sky, the apparent motions of fixed stars, and the position of spectral lines as a function of distance, will probably have come far enough for us to be able to decide empirically the question of whether or not λ vanishes. Conviction is a good mainspring, but a bad judge!"*[7]. More famously, he declared in 1919: *"But this view of the universe necessitated an extension of equations (1), with the introduction of a new universal constant standing in a fixed relation to the total mass of the universe...... This is gravely detrimental to the formal beauty of the theory"*[8].

In July 1917, de Sitter noted that Einstein's modified field equations (2) allowed an alternate cosmic solution, namely the case of a universe with no matter content[9]. (In this model, Einstein's matter-filled three-dimensional universe of closed spatial geometry was replaced by an empty four-dimensional universe of closed *spacetime* geometry.) Einstein was greatly perturbed by de Sitter's cosmology as it was in direct conflict with his understanding of Mach's Principle in these years[4]. He made his criticisms public in a paper of 1918, in which he also suggested that de Sitter's model contained a spacetime singularity[10]. After an intervention by Felix Klein, Einstein privately accepted that the latter criticism was unjustified[11]; however, he never formally retracted his criticism and it is clear from his publications around this time that he did not consider the de Sitter solution a realistic model of the universe [12] [13] [14]. Indeed, we have suggested elsewhere that the existence of a mathematically viable non-Machian solution to the modified field equations for the case of the cosmos may have marked the beginning of Einstein's distrust of the cosmological constant term[15]. It is interesting to note that, while de Sitter did not share Einstein's distrust of a non-Machian universe, he too expressed early reservations about the cosmological constant, declaring: *"It cannot be denied that the introduction of this constant detracts from the symmetry and elegance of Einstein's original theory, one of whose chief attractions was that it explained so much without introducing any new hypothesis or empirical constant"*[16].

## 3. Einstein's rejection of the cosmological constant term

In 1929, the American astronomer Edwin Hubble published the first evidence of a linear relation between the redshifts of the spiral nebulae and their radial distance[17]. Many theorists interpreted the phenomenon as evidence of an expansion of space on the largest scales, and attention turned to non-static relativistic models of the universe that had been proposed in the 1920s by Alexander Friedman[18] and Georges Lemaître[19]. Soon, a number of papers emerged that explored expanding models of the Friedman-Lemaître type for diverse values of cosmic parameters[20].

Einstein was one of the first to accept Hubble's observations as likely evidence of a non-static universe, as evidenced by several statements he made during a sojourn in California in early 1931 (Figure 1). For example, the *New York Times* reported Einstein as commenting that *"New observations by Hubble and Humason concerning the redshift of light in distant nebulae make the presumption near that the general structure of the universe is not static"* [21]. In April 1931, Einstein published a model of the expanding cosmos based on Friedman's 1922 analysis of a matter-filled dynamic universe of positive spatial curvature[22]. The most important feature of this model, sometimes known as the Friedman-Einstein model, was that Einstein formally dispensed with the cosmological constant term for two stated reasons. First, the term was unnecessary because the assumption of stasis was no longer justified by observation: *"Now that it has become clear from Hubbel's [sic] results that the extra-galactic nebulae are uniformly distributed throughout space and are in dilatory motion (at least if their systematic redshifts are to be interpreted as Doppler effects), assumption (2) concerning the static nature of space has no longer any justification"*. Second, the term did not in any case provide a stable solution for a static cosmic radius $P$: *"From these* [Friedman] *equations, one obtains my previous solution by assuming that P is constant over time. However, it can also be shown with* *the help of these equations that this solution is not stable, i.e., a solution that deviates only slightly from the static solution at a given point in time will differ ever more from it with the passage of time. On these grounds alone, I am no longer inclined to ascribe a physical meaning to my former solution, quite apart from Hubbel's* [sic] *observations. Under these circumstances, one must ask whether one can account for the facts without the introduction of the λ-term, which is in any case theoretically unsatisfactory"*[22]. With the cosmological constant term removed, Einstein derived simple expressions relating the rate of cosmic expansion to key parameters such as the present radius of the cosmos, the mean density of matter and the timespan of the expansion. Using Hubble's empirical estimate of the recession rate of the nebulae, he then calculated numerical values for each of these parameters respectively. We have given a first English translation of this paper elsewhere[23] and noted that these calculations contain a slight systematic numerical error; this did not affect a major puzzle raised by the model, namely that the timespan of cosmic expansion implied by the analysis was strangely small in comparison with estimates of the age of the stars and the earth (it was later discovered that the source of the error lay in Hubble's estimates of cosmic distance). Einstein attributed this paradox to the idealized assumptions of the model, in particular the assumption of a homogeneous distribution of matter at early epochs. The paper concludes by once again declaring the cosmological constant term redundant: *"It seems that the general theory of relativity can account for Hubbel's* [sic] *new facts more naturally (namely, without the λ-term), than it can the postulate of the quasi-static nature of space, which has now been rendered a remote possibility empirically"*[22].

In 1932, Einstein collaborated with de Sitter to propose an even simpler model of the expanding universe. Following an observation by Otto Heckmann that the presence of matter in a non-static universe did not necessarily imply a positive curvature of space, and mindful of a lack of empirical evidence for spatial curvature, Einstein and de Sitter set both the cosmological constant and spatial curvature to zero[24]. In this model, the fractional rate of expansion of cosmic radius ($R'/R$) was related to the mean density of matter $\rho$ according to

$$\left(\frac{R'}{R}\right)^2 = \frac{1}{3}\kappa\rho c^2 \tag{4}$$

where $\kappa$ is the Einstein constant. However, the authors were careful not to dismiss the possibility of spatial curvature: *"We must conclude that at the present time it is possible to represent the facts without assuming a curvature of three-dimensional space. The curvature is, however, essentially determinable, and an increase in the precision of the data derived from observations will enable us in the future to fix its sign and to determine its value"*. No such courtesy was afforded to the cosmological constant. Instead, the term was once again declared redundant: *"Historically, the term containing the 'cosmological constant' Λ was introduced into the field equations in order to enable us to account theoretically for the existence of a finite mean density in a static universe. It now appears that in the dynamical case this end can be reached without the introduction of Λ"*[24].

Thus, it is clear that, presented with empirical evidence for a non-static universe, Einstein lost little time in banishing the cosmological constant term from the field equations. Indeed, an early portend of this strategy can be found in a note written by Einstein to Hermann Weyl in 1923. In the course of a discussion of the de Sitter model, Einstein wrote: *"If there is no quasi-static world after all, then away with the cosmological term"* [25] [26]. Taken together, Einstein's removal of the cosmological constant in the Friedman-Einstein model, followed by the removal of spatial curvature in the Einstein-de Sitter model, suggests a minimalist 'Occam's razor' approach to cosmology. Where theorists such as Alexander Friedman, Otto Heckmann and Howard Percy Robertson considered all possible hypothetical universes, it is clear that Einstein's goal was to establish the simplest mathematical model of the universe that could account for observation. In this context, it no surprise that he came to view his introduction of the cosmological constant term as both unnecessary and undesirable.

Many of Einstein's contemporaries took a different approach at this time. Some felt that the cosmological constant term should be retained for reasons of mathematical generality[27 28]; others felt that the term could be used to address cosmological puzzles such as the timespan of cosmic expansion (above) and the question of the formation of galaxies in an expanding universe[29 30]. Some also felt that the term had an important role to play in giving a physical *cause* for cosmic expansion[31 32]. However, Einstein was not swayed by any of these arguments[15].

### 3.1 Einstein's view of the cosmological constant in later years

There is no evidence that Einstein changed his view on the matter in later years. For example, in a substantive but little-known review of relativistic cosmology written for a French publication in 1933, Einstein once again dismissed the cosmological constant term on the twin grounds that it was not necessary for expanding solutions and that it did not in any case give a stable static solution[33]. In 1945, in a review of cosmology for the third edition of *The Meaning of Relativity*, Einstein commented: *"If Hubble's expansion had been discovered at the time of the creation of the general theory of relativity, the cosmologic member would never have been introduced. It seems now so much less justified to introduce such a member into the field equations, since its introduction loses its sole original justification"*[34]. In 1947, Einstein explained in a letter to Lemaître why he found the term so unappealing: *"Since I have introduced this term I always had a bad conscience. But at that time I could see no other possibility to deal with the fact of the existence of a finite mean density of matter. I found it very ugly indeed that the field law of gravitation should be composed of two logically independent terms which are connected by addition. About the justification of such feelings concerning logical simplicity it is difficult to argue. I cannot help but feel it strongly and I am* *unable to believe that such an ugly thing should be realised in nature*[35]. This statement can be read in context in Figure 2. Finally, in the famous Einstein Festschrift of 1949, Einstein provided his last formal response to arguments in favour of retaining the term:

> As concerns Lemaître's arguments in favour of the so-called "cosmological constant" in the equations of gravitation, I must admit that these arguments do not appear to me as sufficiently convincing in view of the present state of our knowledge. The introduction of such a constant implies a considerable renunciation of the logical simplicity of theory, a renunciation which appeared to me unavoidable only so long as one had no reason to doubt the essentially static nature of space. After Hubble's discovery of the "expansion" of the stellar system, and since Friedmann's discovery that the unsupplemented equations involve the possibility of the existence of an average (positive) density of matter in an expanding universe, the introduction of such a constant appears to me, from the theoretical standpoint, at present unjustified.[36]

Thus there is little doubt that, from the 1930s onwards, Einstein had little use for the cosmological constant term on the dual grounds that it did not give a stable static solution, and that the term was not required in order to describe the dynamic universe suggested by observation. Einstein's emphasis on the first of these points in all of his later writings on cosmology is of primary interest here as it implies that he came to view his failure to note the instability of his 1917 model as a technical error. Indeed, it could be argued that the error prevented the prediction of a dynamic cosmos a decade before Hubble's observations.

Over the years 1950-1990, many other physicists came to share Einstein's view of the cosmological constant as a blunder. One reason was the resolution of the so-called age paradox (above); with improved measurements in the 1950s and 60s of the distance to the galaxies, estimates of the age of the universe implied by the Hubble expansion were no longer in conflict with the age of the stars[15]. Confidence in the simplest models of the cosmos increased and many physicists came to view the cosmic constant term as an unnecessary complication. Indeed, in the absence of any evidence for spatial curvature or for a cosmological constant, the Einstein-de Sitter model (above) became the standard cosmic model for astronomers and theoreticians. For many years, it seemed that the cosmos might be described in terms of just two parameters, each of which could be determined independently by astronomy, a view that remained essentially unchanged until the emergence of the first evidence for an accelerated expansion in the late 1990s.[15]

## 4. The legend of Einstein's "biggest blunder"

In a substantial article on 'big bang' cosmology published in *Scientific American* in May 1956, the Russian émigré physicist George Gamow reported: *"Einstein remarked to me many years ago that the cosmic repulsion idea was the biggest blunder he had made in his entire life"*[37]. We note first that the term biggest blunder is not quoted in parenthesis; thus it is not clear whether Gamow claims that Einstein used these exact words. Indeed, it is not stated whether the conversation was conducted in German or English. That said, Gamow made a very similar report some years later, in his autobiography of 1970[38]:

> Studying Einstein's publications on that subject from a purely mathematical point of view, Friedmann noticed that Einstein had made a mistake in the alleged proof that the universe must necessarily be stable and unchangeable in time….Einstein's proof does not hold and Friedmann realised that this opened up an entire world of time-dependent universes: expanding, collapsing, and pulsating ones. Thus, Einstein's original gravity equation was correct, and changing it was a mistake. Much later, when I was discussing cosmological problems with Einstein, he remarked that the introduction of the cosmological term was the biggest blunder he ever made in his life.

In time, Gamow's report of Einstein's "biggest blunder" remark became one of the great stories of 20[th] century physics. This is hardly surprising, given Einstein's enormous profile in the international media as the epitome of genius; that such an individual saw himself as a blunderer who made many mistakes, including one of cosmic proportions, was great copy. Another reason may have been that the story reflected the view of the community: as noted in section 3, a great many physicists and astronomers came to share Einstein's rejection of the term in the years 1950-1990. That said, it is worth noting that the cosmological constant term was re-introduced to cosmology on various occasions in order to address outstanding puzzles in astronomical observation as they arose, culminating in today's observation of an accelerated cosmic expansion.[15] Each time this occurred, the tale of Einstein's "biggest blunder" was retold in the scientific literature and in the popular scientific press, strengthening the legend.

## 5. Interrogating the legend: the case for the prosecution

In recent years, Einstein's reputed "biggest blunder" remark has become the subject of increasing scepticism. At first, this suspicion took the form of mild doubts expressed by the occasional physicist or historian [39] [40] [26]. More recently, these doubts have coalesced into a widespread scepticism. In 2013, a painstaking historical study by the noted astronomer Mario Livio concluded that Einstein probably never made the statement[41]. Today, it is rare to encounter a reference to Einstein's "biggest blunder" remark that is not qualified by a disclaimer to the effect that the statement may be an invention on the part of Gamow [42] [43] [44]. Different doubts have been expressed by different authors, but they can be grouped into three effective classes: the question of primary sources, the question of Gamow's character, and the question of the interaction between Einstein and Gamow.

*(i) The question of primary sources*

It comes as something of a shock that no record of Einstein's famous remark has been found in his own scientific papers, personal papers, letters or books. Until recently, almost all known references to the comment could be traced back to a single source – George Gamow. Historians mark a clear distinction between primary and secondary sources, and thus the lack of a primary source for Einstein's remark raises serious doubts concerning the accuracy of the story[40][41][42][44]. After all, Einstein committed a great deal to print during his lifetime: is it not strange that he did not put this famous remark into print at least once?

Worse, a contemporaneous record *does* exist of Einstein claiming to have made one great mistake, but in a very different context. Describing a meeting with Einstein in 1954, the Nobel laureate Linus Pauling famously recorded in his diary: *"He said that he had made one great mistake – when he signed the letter to President Roosevelt recommending that atom bombs be made"*[45]. As Livio has pointed out[41], Pauling's record would seem to cast doubt on our legend, as a scientist can presumably have only one biggest mistake.

*(ii) The question of Gamow's character*

Another problem concerns the reliability of our secondary source. It is known from Gamow's own writings and from the accounts of colleagues that he was a physicist with an unusually well-developed sense of humour, given to pranks and to hyperbole[38] **Error! Bookmark not defined. Error! Bookmark not defined.**. These pranks were not always left at the office door. Several august bodies, such as the US National Academy of Sciences and the journal *Naturwissenschaften* (the German equivalent of *Nature*) fell victim to Gamow's hoaxes and humour. In one famous episode, Gamow added the name of renowned stellar physicist Hans Bethe to a key article on nucleosynthesis in the early universe written with his postgraduate student Ralph Alpher, so that the paper's author list would reflect its primordial theme[46]. More seriously, Gamow had a well-earned reputation as a drinker and bon viveur, and struggled with alcoholism later in life [47] [48]. It seems unfortunate that, of all Einstein's acquaintances, the burden of proof should fall on such an apparently unreliable witness.

*(iii) The interaction between Einstein and Gamow*

A related problem concerns the interaction between Einstein and Gamow. In his autobiography of 1970, Gamow presents a pleasant narrative of regular meetings with Einstein in the course of their work as consultants for the Bureau of Ordnance of the American Navy[38]:

> A more interesting activity during that time was my periodic contact with Albert Einstein, who ….served as a consultant for the High Explosive Division. Accepting this consultantship, Einstein stated that because of his advanced age, he would be unable to travel periodically from Princeton to Washington D.C. and back, and that somebody must come to his home in Princeton, bringing the problems with him. Since I happened to have known Einstein earlier, on non-military grounds, I was selected to carry out this job. Thus on every other Friday I took a morning train to Princeton carrying a briefcase tightly packed with confidential and secret Navy projects…
>
> After the business part of the visit was over, we had lunch either at Einstein's home or at the cafeteria of the Institute for Advanced Study, which was not far away, and the conversation would turn to the problems of astrophysics and cosmology. In Einstein's study, there were always many sheets of paper scattered over his desk and on a nearby table, and I saw that they were covered with tensor formulae which seemed to pertain to the unified-field theory, but Einstein never spoke about that. However, in discussing purely physical and astronomical problems he was very refreshing, and his mind was as sharp as ever.

However, as recently noted by Livio[41], a 1986 article by Stephan Brunauer, the physicist and US Navy lieutenant who recruited both Einstein and Gamow as consultants for the Navy, presents a different account[49]:

> Gamow, in later years, gave the impression that he was the Navy's liaison man with Einstein, that he visited every two weeks, and the Professor 'listened' but made no contribution – all false. The greatest frequency of visits was mine, and that was about every two months.

This account clearly suggests that Gamow may have exaggerated the nature and number of his meetings with Einstein on Navy matters, raising the possibility that Gamow's famous report of Einstein's remark was merely one incident in a process of exaggeration and self-aggrandization. Livio also points out that there is little correspondence between Einstein and Gamow to be found in Einstein's papers – thus it seems surprising that Einstein would use an informal expression such as "biggest blunder" in conversation with Gamow and not with any of his more intimate friends and colleagues[41].

## 6. Interrogating the legend - the case for the defence

We find many of the arguments above quite reasonable. Yet, as noted in section 3, we also find Einstein's reported remark highly compatible with his attitude towards the cosmological constant term in the years after 1930. Indeed, it is worth repeating that, in all of his later reviews of cosmology [22] [33] [34], Einstein stressed the instability of his static model of 1917 as grounds for rejection of the model (in addition to Hubble's observations). Thus, there is little question that Einstein came to view his introduction of the cosmological constant term as a serious error and we find it worthwhile to revisit the arguments of section 5.

*(i) On the question of primary sources*

There is no question that the lack of a primary record of Einstein's famous remark in his own papers raises some doubts about the story. However, several caveats pertaining to this point should be noted. In the first instance, it is unlikely that Einstein would have included such an informal comment in any of his formal scientific papers. Like all scientific pioneers, Einstein made many technical errors in his work; however, he rarely formally corrected or commented upon such errors afterwards. For example, Einstein's statement in a paper of 1918[10] that the de Sitter model contained a singularity was not correct; although he later conceded this point in a letter to Felix Klein, he never amended his paper on the subject (see section 2). There are many other examples of mistakes by Einstein that were never formally corrected or commented upon in the literature, from calculations of the viscosity of fluids in 1905 to early proofs of the relation $E = mc^2$. [50]

As regards Einstein's personal papers, it is indeed a little surprising that the "biggest blunder" remark is not to be found in his later correspondence with old friends such as Erwin Schrödinger, Wolfgang Pauli or Kornel Lanczos. However, it should be noted that almost all of Einstein's correspondence of a scientific nature in these years is concerned with his search for a unified field theory or his views on the interpretation of quantum physics. As will be discussed below, Einstein appears to have had very little interaction with physicists interested in cosmology, with Gamow as a notable exception.

A third caveat on the question of primary sources is that the personal papers and letters written by Einstein in later years have not yet been subject to the same scrutiny as his earlier papers. Many readers will be aware of the Collected Papers of Albert Einstein (CPAE), a unique initiative that has made Einstein's scientific and personal papers publicly available online in both German and English, with accompanying editorial notes. However, this project extends only to the year 1928 so far. For materials beyond this point, scholars rely on resources such as the Albert Einstein Archive of the Hebrew University of Jerusalem. Many of the materials at such archives are available on request only and are not available in translation. In particular, it is not yet possible to search the handwritten manuscripts for a particular statement. In addition, we are informed that some of Einstein's last letters have not yet been collated and catalogued.[a] Thus, when it is stated that no record of Einstein's statement exists in his personal papers, it would perhaps be more accurate to state that no record of the statement has been found in Einstein's papers to date.

Considering Einstein's reported use of the words *"one great mistake"* when discussing his letter to President Roosevelt with Pauling (see above), we note that this was clearly a reference to an error that was *political*, rather than *scientific,* in nature. Thus we find the point rather moot - one could have a greatest scientific blunder as well as a greatest political blunder. We note in passing that this particular statement by Einstein is widely accepted without question, although no record of the statement has been found in Einstein's papers. Indeed, the only supporting evidence for the statement stems from a colleague's recollections of a conversation with Einstein - exactly as in the case of the "biggest blunder" remark.

*(ii) On the question of Gamow's character*

As regards the character of George Gamow, one wonders whether too much has been made of Gamow's famous sense of humour. It seems to us rather a stretch to assume that, because Gamow was given to pranks, he may have fabricated a key comment on cosmology by Einstein. For example, it's worth reconsidering the story of Gamow's inclusion of Bethe's name as co-author of the famous αβγ paper (above). In the first instance, Gamow was on very friendly terms with Bethe, a world-famous authority on stellar processes, and the two had many interactions in the field of nuclear physics; indeed, Bethe was an early and enthusiastic contributor to the well-regarded annual conference on theoretical physics hosted by Gamow at George Washington University.[51] Second, it is known, but seldom acknowledged, that the αβγ paper was reviewed and approved by Bethe before publication [52] [53]. Finally, we note that Bethe acted as external examiner for Alpher's doctoral thesis just a few months later. Thus, the inclusion of Bethe's name as a co-author on a key paper may have been a clever pun, but it was hardly the mischievous, random act that is customarily portrayed.[54]

---

[a] We thank Daniel Kennefick for this information.

Gamow's reputation for hyperbole and japes certainly gives pause for thought. However, one wonders to what extent the perception of, and reaction to, his behaviour arose from cultural differences. As pointed out by the Nobel laureate James Watson, what seemed outrageous to some American scientists in the 1940s might have attracted less comment in Europe or in Russia[47]. There is little question that Gamow's drinking became a serious problem in his later years; however, we have found no records of an untoward effect on his work in the 1940s and 50s. Indeed, one wonders to what extent Gamow's own tall tales of his exploits may have influenced the later perception of his reliability. As the noted astronomer Vera Rubin, a former postgraduate student of Gamow's noted: *"It is true that Gamow was funny and that he drank. It is also true that he was a brilliant scientist, devoted friend and concerned teacher, whose intuition exceeded that of any scientist I have known"*[55].

*(iii) On the interactions between Einstein and Gamow*

As regards Gamow's interactions with Einstein, Stephan Brunauer's account that Gamow may have exaggerated his role in liaising with Einstein on behalf of the US Navy seems persuasive. However, a passage from the memoir of well-known American physicist Robert Finkelstein gives pause for thought[56]:

> At about that time Einstein had agreed to serve as a consultant to our group but did not want to travel to Washington. So there had to be a liaison person and I was given that opportunity. Since Einstein did not know me, there had to be someone to introduce us. It then happened that I was introduced to Einstein by John von Neumann, one of the most important mathematicians of all time, and who had also become a consultant to our group. It was a very great experience for a new Ph.D. to be introduced to Einstein by Von Neumann! During the following period I met Einstein every week until Gamow joined our group and became the liaison person.

It seems reasonable to conclude from this account that Gamow, as Finkelstein's successor in the role of Einstein's 'liaison person' with the Navy, also met Einstein regularly, at least in the early months of his appointment. Thus, one wonders if the truth lies somewhere in between Brunaeur's recollections and Gamow's comments in his autobiography.

It is also important to note that Gamow's interaction with Einstein was not limited to their work as consultants for the US Navy. It is known that Gamow visited Princeton on many occasions in the 1940s and 1950s [51] [57]. Many biographers have noted that Einstein became quite isolated from scientific colleagues at Princeton in later years [58] [59] [60] and it is likely that Gamow, a German-speaking physicist with a reputation as an 'ideas' man, was a welcome visitor. Gamow never pretended to any great expertise in the technicalities of mathematical physics, but his extraordinary insights into nuclear and quantum physics led to a number of successful breakthroughs in different areas. For example, Gamow's early grasp of Schrödinger's wave mechanics  led to the first successful explanation of the alpha-decay of the nucleus in 1928 [51] [61], a spectacular triumph for the young theorist that propelled him into the front rank of international physics (Figure 3). Like all leading physicists, Einstein would have been aware of this success; indeed, it is likely that he received first-hand reports of this work from old friends such as Max Born and Niels Bohr.[b] In a second important advance, Gamow persuaded Ernest Rutherford's group at the Cavendish Laboratory at Cambridge that the phenomenon of quantum tunneling might allow the penetration of the atomic nucleus by particles at relatively low energy (Figure 4), a suggestion that led directly to the famous splitting of the atomic nucleus by John Cockcroft and Ernest Walton in 1932[54] [51] [62]. Indeed, this experiment was much appreciated by Einstein as the first experimental verification of $E = mc^2$ [63] [64].Later in the 1930s, Gamow's knowledge of nuclear and quantum physics played an important role in the development of the Bohr-Gamow 'liquid–drop' model of the nucleus [51] [61], while in the 1940s, the pioneering research of Gamow and his colleagues into nuclear physics in the early universe laid the foundations of the modern theory of primordial nucleosynthesis [65].[c]

The latter work is the most relevant here. From the mid-1940s onwards, Gamow became interested in the question of whether the observed abundance of the chemical elements might be explained in the context of the expanding cosmologies of Friedman and Lemaître; could it be that the elements were formed, not in the cauldron of stars as commonly supposed, but in a universe that was once extremely hot and dense? Shortly after completing a key paper on the subject [66], Gamow sent a copy to Einstein. The latter responded positively [67]stating:

> After receiving your manuscript I read it immediately and then forwarded it to Dr. Spitzer. I am convinced that the abundance of the elements as function of the atomic weight is a highly important starting point for cosmogenic speculations. The idea that the whole expansion process started with a neutron gas seems to be quite natural too. The explanation of the abundance curve by formation of the heavier elements in making use of the known facts of probability coefficients seems to me to be pretty convincing.

[b]Gamow studied with Born's group in Göttingen in 1928 and with Bohr's group in Copenhagen in 1929-31.
[c] We note that Gamow was nominated for the Nobel Prize in physics on at least three occasions (1943, 1946, 1967).

This is a very favourable review and it also confirms that Einstein and Gamow were discussing astronomy and cosmology in the late 1940s, at a time when Einstein had very little interaction with other physicists and virtually no interaction with physicists interested in cosmology. In this context, one might argue that, if Einstein was to make the "biggest blunder" comment to anyone, Gamow was a very likely candidate. A copy of Einstein's letter forwarded by Gamow to a colleague is displayed in Figure 5; careful perusal shows that it is dated 1948, not 1946 as sometimes stated.[d] We also note that the letter suggests that Einstein and Gamow interacted through English rather than German. Finally, we note that Gamow added the inscription *"Of course, the old man agrees with almost anything nowaday"* at the bottom of the letter. This inscription is sometimes interpreted in a negative way[41]; however, we find it a good example of the manner in which Gamow often undermined his own successes with humour and self-deprecation.

Thus it seems very probable that Einstein and Gamow enjoyed many interesting conversations in the 1940s and 50s on a wide range of physics, just as Gamow reported in his autobiography. Which of these conversations occurred under the auspices of the US Navy is rather a moot point, as Gamow does not claim that Einstein's remark was made during one of their Navy meetings.

*(iv) A new line of evidence*

We note finally that the story of Einstein's "biggest blunder" does not in fact rest entirely on Gamow's testimony alone. In the course of our own research into Einstein's cosmology in his later years, we have come across similar testimonies from two other physicists.

---

[d] We thank Orith Burla Barnea of the Albert Einstein Archive for confirming this.

In the book *Exploring Black Holes: Introduction to General Relativity*, the highly respected theorist John Archibald Wheeler states: *"Going into the doorway of the Institute for Advanced Study's Fuld Hall with Einstein and George Gamow, I heard Einstein say to Gamow about the cosmological constant, "That was my biggest blunder of my life"* [68]. This report can be read in context in Figure 6.

Something similar was reported by Gamow's erstwhile colleague Ralph Alpher in 1998. In response to a query from the noted historian of astronomy Joseph S. Tenn on the History of Astronomy Listserve (HASTRO-L), an online discussion forum used by physicists, astronomers and historians, Alpher recalled a meeting with Einstein at Princeton at which the problem of the timespan of cosmic expansion was discussed. Alpher's recollection was that Einstein described his original introduction of the term as a blunder: *"A way to fix this was to reactivate the cosmological constant. Einstein did not like this very much, and, as I recall, said his introduction of the concept in his early work was a blunder"* [69]. This account can be read in full in Figure 7. It confirms our earlier impression that, even in the face of the problematic timespan of cosmic expansion, Einstein saw the use of the cosmological constant term in his later years as an error. It is of course possible that both Wheeler and Alpher were influenced by Gamow's recollections. However, it seems a stretch to accuse three different scientists of invention; a more likely explanation is that all three reports pertain to the same occasion.

## 7. Summary and conclusions

Summarizing the results of our interrogation into the legend of Einstein's "biggest blunder" remark, we find that:

(i)     The statement is highly compatible with Einstein's cosmic models of the 1930s

(ii)    The statement is highly compatible with Einstein's later writings on cosmology

(iii)    No evidence of the statement is to be found in Einstein's formal scientific works or in his personal papers; however, we would not expect the former, while the latter is not yet established beyond doubt

(iv)    It is very plausible that Einstein made the remark to George Gamow in particular, as the latter was a scientist whose work he respected and with whom he is known to have discussed astronomy and cosmology

(v)    Two other physicists have reported a similar statement by Einstein

In conclusion, there is little doubt that Einstein came to view his introduction of the cosmological constant term as a serious error, and we find it very likely that he spoke of the term as his "biggest blunder" on at least one occasion. While it is difficult to establish for certain that Einstein used those exact words, we see no reason to doubt the reports of witnesses to the remark.

## Coda

We note finally that it is sometimes argued that Einstein's real blunder was not the introduction of the cosmological constant in 1917, but his banishment of the term from 1931 onwards. As is well-known, modern measurements of cosmic expansion and of the cosmic microwave background suggest the presence of a significant 'dark' component of cosmic energy, a phenomenon that can be described within the context of general relativity by including a cosmological constant term in the field equations [15] [70]. In addition, the term increases the generality of the field equations and may help establish links between general relativity and other modern field theories [71] [72]. However, we find this argument somewhat ahistorical. As pointed out in section 3, Einstein's cosmology was focused on the attempt to describe the observed universe as simply as possible, at a time when the discovery of dark energy and the advent of quantum field theories lay in the distant future. By contrast, Einstein came to view his failure to consider the stability of his static model a true error, and he acknowledged it as such on many occasions.

We shall never know Einstein's reaction to the recent discovery of an acceleration in cosmic expansion. At first sight, it seems likely that he would have been pleased that the phenomenon can be described within the context of the general theory of relativity, albeit via the re-introduction of the cosmological constant term. However, Einstein's clear dislike of the term as a complication of the field equations gives pause for thought; is it possible we are once again falling into the trap of amending the general theory of relativity in an ad-hoc manner in order to account for a cosmological puzzle that may one day be described without recourse to such changes? Thus, Einstein's "biggest blunder" remark provides some succour to those cosmologists today who seek to describe the observed universe without recourse to a cosmological constant term.[73]

## Acknowledgements

The authors wish to acknowledge the use of online materials in the Collected Papers of Albert Einstein (CPAE), an important historical resource published by Princeton University Press in conjunction with the California Institute of Technology and the Hebrew University of Jerusalem. We also thank the Hebrew University of Jerusalem for permission to display the Einstein letters shown in Figure 2 and Figure 5. Cormac O'Raifeartaigh thanks the Dublin Institute for Advanced Studies for the use of research facilities and Norbert Straumann, Werner Nahm and Michael O'Keeffe for helpful discussions. Simon Mitton thanks St Edmund's College, University of Cambridge for the support of his research in the history of science.

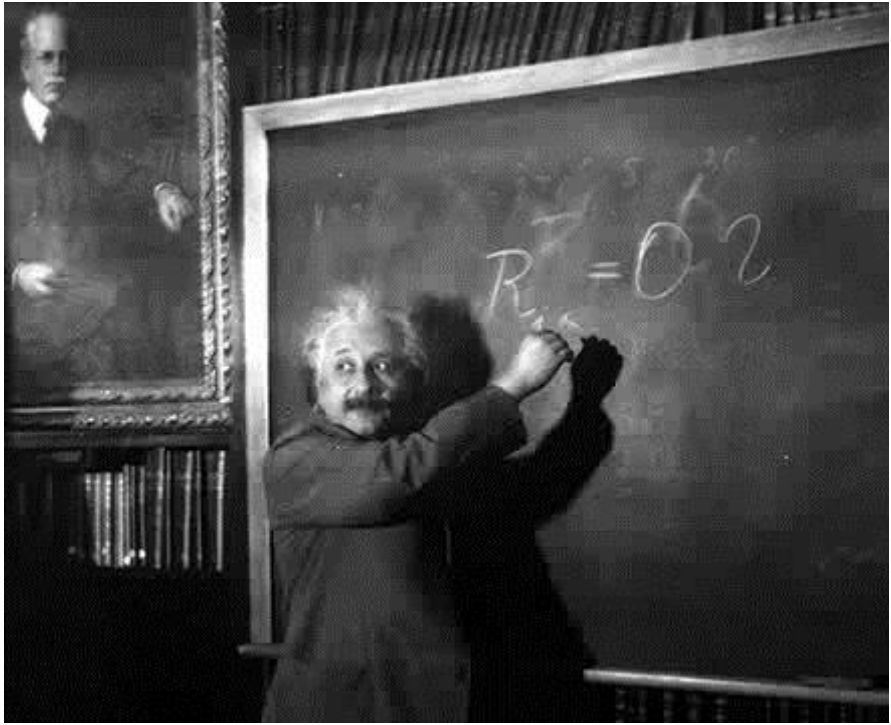

*Figure 1*

Einstein at the blackboard in 1931 in the Hale Library of the Mount Wilson Observatory in

Pasadena, California. ©The Huntington Library, Art Collections.

September 26,1947

Professor G.Lemaitre
9 rue Henry de Braekeleer
Brussels,Belgium

Dear Professor Lemaitre:

I thank you very much for your kind letter of July 30th. In the meantime I received from Professor Schillpp your interesting paper for his book. I doubt that anybody has so carefully studied the cosmological implications of the theory of relativity as you have. I can also understand that in the shortness of $T_0$ there exists a reason to try bold extrapolations and hypotheses to avoid contradiction with facts. It is true that the introduction of the $\Lambda$ term offers a possibility, it may even be that it is the right one.

Since I have introduced this term I had always a bad conscience. But at that time I could see no other possibility to deal with the fact of the existence of a finite mean density of matter. I found it very ugly indeed that the field law of gravitation should be composed of two logically independent terms which are connected by addition. About the justification of such feelings concerning logical simplicity it is difficult to argue. I cannot help to feel it strongly and I am unable to believe that such an ugly thing should be realized in nature.

*Figure 2*

Letter from Einstein to Georges Lemaître, September 26th, 1947. © The Hebrew University of Jerusalem.

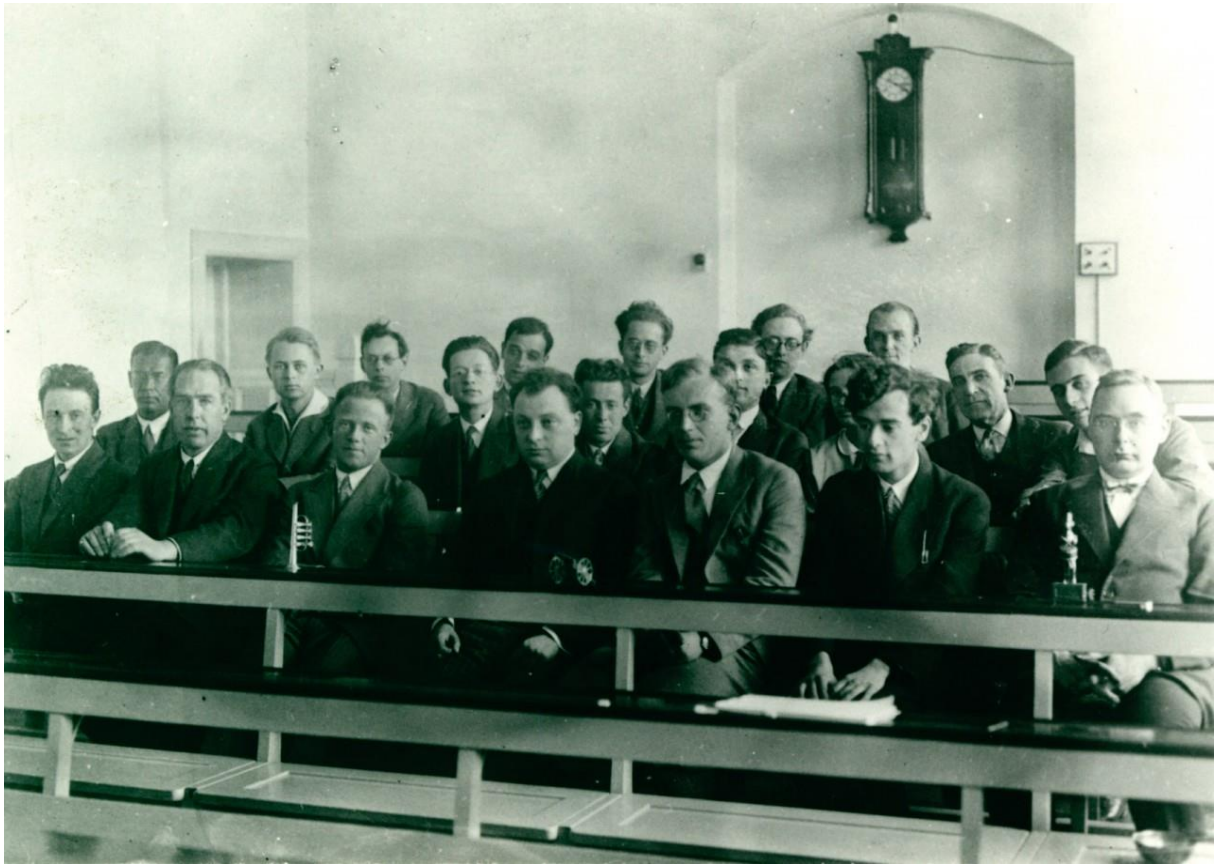

*Figure 3*

George Gamow (aged 26) at a meeting in 1930 at the Niels Bohr Institute in Copenhagen.

Front row (L to R): Oskar Klein, Niels Bohr, Werner Heisenberg, Wolfgang Pauli, Gamow,

Lev Landau and Hans Kramers. © Niels Bohr Archive, Copenhagen.

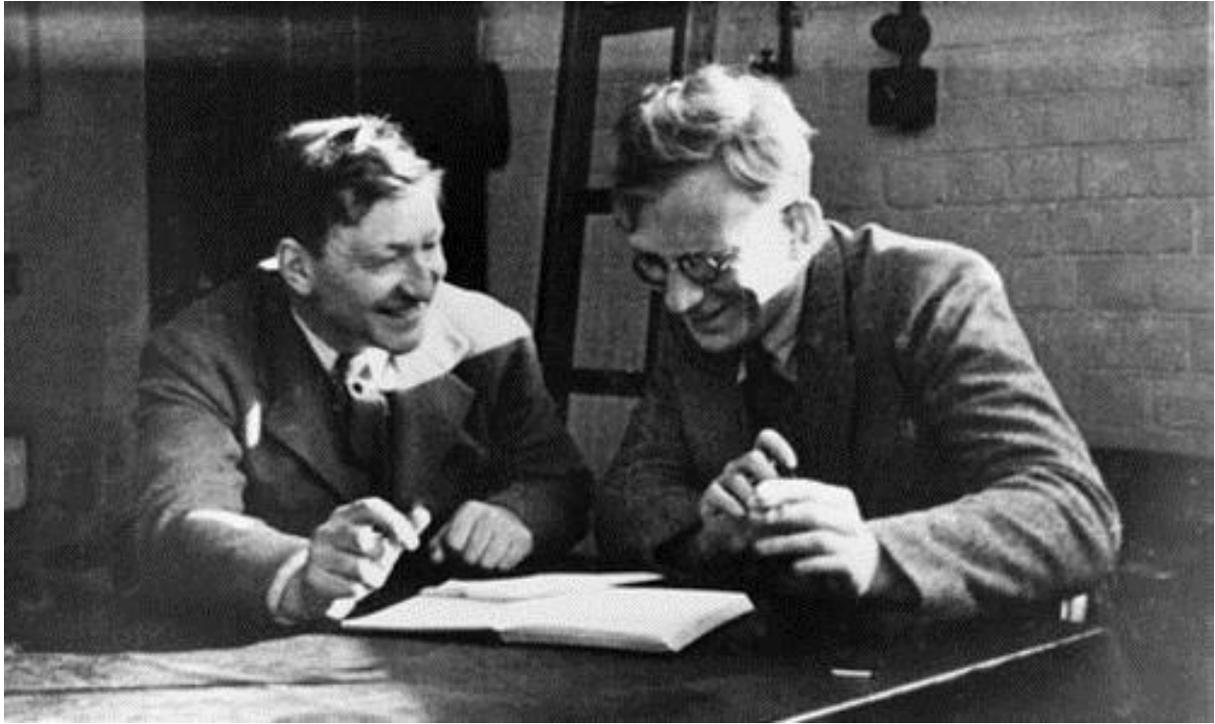

*Figure 4*

John Cockcroft (L) and George Gamow (R) at the Cavendish Laboratory in 1931, discussing the possibility of splitting the atomic nucleus. © Cavendish Laboratory Photography Collection.

August 4,1948

Professor G.Gamov
Ohio State University
Columbus,Ohio

Dear Mr.Gamov:

After receiving your manuscript I read it immediately and then forwarded it to Dr.Spitzer. I am convinced that the abundance of elements as function of the atomic weight is a highly important starting point for cosmogonic speculations. The idea that the whole expansion process started with a neutron gas seems to be quite natural too. The explanation of the abundance curve by formation of the heavier elements in making use of the known facts of probability coefficients seems to me pretty convincing. Your remarks concerning the formation of the big units (nebulae) I am not able to judge for lack of special knowledge.

Thanking you for your kindness,I am

yours sincerely,

A. Einstein.

Albert Einstein.

*Of course, the old man agrees with almost everything nowadays. Geo.*

*Thanks for slides G.*

*Figure 5*

Letter from Einstein to George Gamow, August 4th 1948, with hand-written inscription added by Gamow. © The Hebrew University of Jerusalem.

**Opinion: The Bang-to-Crunch Universe
Too Simple to be Wrong!**

*Spacetime tells mass how to move, and mass tells spacetime how to curve.* If the black hole provides our closest plain-speaking witness to spacetime curvature, the Cosmos itself looks like the one operative on the largest scale of space and time. The stars bear witness to the scales of space and time characteristic of the Cosmos. Sun gets its energy by burning hydrogen to helium and some of that helium to heavier elements, profiting from the difference in mass per nucleon between hydrogen and helium: 1.00783 for hydrogen and 1.00065 for helium (in units that set the mass of the most common isotope of carbon equal to 12).

That space is expanding shows most directly in the red shift of light from distant galaxies, a red shift which is greater the more distant the galaxy. The inflation of a balloon (Figure 1) provides a simple model for such an expansion of the Universe. This model tells its story in the functional dependence of the radius $R(t)$ of the sphere on the time $t$. In this model the curvature of spacetime in the large has only two components. One is the momentary "intrinsic curvature," fixed by the momentary radius $R(t)$ of the idealized three-geometry: $6/R^2(t)$. The other contribution to the curvature arises from the variation of this radius with the cosmological time, $t$: $(6/R^2)(dR/dt)^2$. Einstein, considering the matter in the days before Hubble had seen and measured the expansion of the Universe, found it natural to think of a closed Universe with an essentially constant radius $R$.

At this point one comes hard up against the second part of gravitation theory: mass tells spacetime how to curve. In Einstein's time there did not seem to be enough mass around to curve up the Universe into closure. Therefore Einstein postulated an additional source of curvature, a so-called "cosmological constant." Going into the doorway of the Institute for Advanced Study's Fuld Hall with Einstein and George Gamow, I heard Einstein say to Gamow about the cosmological constant, "That was the biggest blunder of my life." If we drop that term, then the equation in which matter tells spacetime how to curve becomes

$$\left(\frac{1}{R}\frac{dR}{dt}\right)^2 + \frac{1}{R^2} = \frac{8\pi}{3}\rho$$

This equation forecasts the connection between radius and time depicted by the cycloid in Figure 4 (explanation in the caption to that figure). The Universe begins with a Big Bang. In a phase of gradually slowing expansion, it reaches a maximum radius and recontracts to zero radius—a "big bang" to "big crunch" history.

An article by John Noble Wilford in the Science Times section of the *New York Times* for Tuesday March 3, 1998, reports observations by two separate groups of investigators which they interpret as showing that today the expansion of the Universe is speeding up rather than undergoing the slowdown expected for any approach to maximum expansion. [See box page G-20.] Later that day I encountered a hard-bitten veteran gravitation physics colleague in the elevator of the Princeton physics building and asked him if he believed the purported evidence of accelerating expansion. "No," he replied. Neither do I. Why not? Two reasons: (1) Because the speed-up argument relies too trustingly on the supernovas being standard candles. (2) Because such an expansion would, it seems to me, contradict a view of cosmology too simple to be wrong. Such clashes between theory and experiment have often triggered decisive advances in physics. We can hope that some decisive advance is in the offing.

— John Archibald Wheeler

*Figure 6*

Excerpt from the book *Exploring Black Holes: Introduction to General Relativity* by Edwin Taylor and John Wheeler, in which Wheeler recalls Einstein describing the cosmological constant as his *"biggest blunder"*. © Addison Wesley Longman.

===================================================

George Gamow, Bob Herman and I had a visit with Einstein at Princeton in, if my memory serves me correctly, about 1952. Among items we discussed was the problem of the age of the standard Big Bang model (not what we called it then). We had been using an expansion rate put forward by a German astronomer named Behr. His work did not survive for very long, but it did give us some heartburn as it led to an age of the universe less than the age of the earth, and it impressed steady state cosmologists at the time as strong evidence against the Big Bang. A way to fix this was to reactivate the cosmological constant. Einstein did not like this very much, and, as I recall, said his introduction of the concept in his early work was a blunder. The only documentation I know of was Gamow's remark in his autobiography.

*Figure 7*

Posting by Ralph Alpher on the online message board of the History of Astronomy Discussion Group (HASTRO) on April 2nd 1998.